\documentstyle[12pt]{article}
\begin{document}
\title{{\bf SENSIBLE QUANTUM MECHANICS:\\ARE PROBABILITIES ONLY IN THE MIND?}
\thanks{Alberta-Thy-13-95, gr-qc/9507024, to be published in {\em Proceedings
of the Sixth Seminar on Quantum Gravity, June 12-19, 1987, Moscow, Russia},
edited by V. A. Berezin and V. A. Rubakov (World Scientific, Singapore, 1996)}}
\author{
Don N. Page
\thanks{Internet address:
don@phys.ualberta.ca}
\\
CIAR Cosmology Program, Institute for Theoretical Physics\\
Department of Physics, University of Alberta\\
Edmonton, Alberta, Canada T6G 2J1
}
\date{(1995 July 4)}

\maketitle
\large
\begin{abstract}
\baselineskip 14.7 pt

Quantum mechanics may be formulated as {\it Sensible Quantum Mechanics} (SQM)
so that it contains nothing probabilistic except conscious perceptions.  Sets
of these perceptions can be deterministically realized with measures given by
expectation values of positive-operator-valued
{\it awareness operators}.  Ratios of the measures for these sets of
perceptions can be interpreted as frequency-type probabilities for many
actually existing sets.  These probabilities generally cannot be given by the
ordinary quantum ``probabilities'' for a single set of alternatives.
{\it Probabilism}, or ascribing probabilities to  unconscious aspects of the
world, may be seen to be an
{\it aesthemamorphic myth}.
\\
\\
\end{abstract}
\normalsize
\baselineskip 14.7 pt

\section{\it Introduction}

\hspace{.25in}Probabilities are the most mysterious aspect of quantum
mechanics, in my mind.

There is first the mystery of which amplitudes should be squared to give
probabilities.  Should it be the amplitudes of ``measurement outcomes''?  If
so, what constitutes a ``measurement''?  Should it be the amplitudes of all
``events,'' whether or not they is ``measured''?  If ``events'' are represented
by projection operators, should one just calculate them for one set that
commutes and adds up to the identity operator?  If so, which such set of
``events'' should one choose?  Should one instead calculate probabilities for
``histories''?  If so, which set of ``histories'' should one choose?

There is second the even deeper mystery of what the resulting probabilities
mean.  One interpretation is that they are ``propensities'' for certain
``possibilities'' to be ``actualized.''  In this interpretation a unique one of
the possible ``measurement outcomes,'' ``events,'' or ``histories'' actually
occurs, with the probability assigned to it by quantum mechanics, and the other
possibilities do not.  But then what chooses the set of ``measurement
outcomes,'' ``events,'' or ``histories''?  For example, if it is the set of
events given by a set of commuting rank-one projection operators onto an
orthonormal basis for the Hilbert space of states, what determines this basis?
Furthermore, once the set is chosen so that the quantum-mechanical
probabilities can be determined from the state, what is it that actually
determines which ``possibility'' from that set is ``actualized''?  If the
probabilities of quantum mechanics are interpreted in this propensity sense as
indicating fundamental uncertainties, then quantum mechanics itself would be
uncertain and incapable of being a complete theory of the universe in which
only certain possibilities are actualized.

The last question above, but not the two preceding it, can be avoided by taking
the alternative interpretation that probabilities are ``frequencies'' for many
``actualities.''  This is a ``many-worlds'' interpretation \cite{E}.  If the
set of possibilities (now all considered to be actualities, at least all those
possibilities with nonzero frequency-type probabilities) were determined, and
if this many-worlds interpretation of quantum mechanics were a true
representation of the universe, then this quantum mechanical theory would be
definite and complete, for it would completely specify the frequency of all
actualities.

Of course, there is the technical problem of interpreting the probabilities as
frequencies of a finite integral number of actualities if any of the
probabilities are irrational numbers, but one can circumvent that difficulty by
interpreting the probabilities as ratios of measures of a continuum of
actualities (though perhaps at the cost of no longer having a clear intuitive
grasp of the concept in terms of familiar objects, but that should be no
fundamental difficulty, other than for one's intuition).

However, there still remains within this many-worlds quantum theory the
unresolved question of what the set of actualities is.  This certainly does not
make the many-worlds interpretation {\it worse} then those with a propensity
interpretation of the probabilities, but since those other interpretations have
the even more basic problem of not saying which possibilities are actualized,
it is often overlooked that they also share with the many-worlds interpretation
the more subtle problem of which set of possibilities (actualities in the
many-worlds interpretation) is picked out.

Here I shall summarize a version of quantum mechanics called {\it Sensible
Quantum Mechanics} (SQM)
[2-6] in which not only are probabilities frequencies (or, more precisely,
ratios of measures of sets) rather than propensities, but also probabilities
apply only to {\it conscious perceptions}.  The unconscious aspect of our
universe (here called the {\it quantum world}) is completely described by the
quantum state and its complex {\it amplitudes} (the {\it expectation values} of
all operators, which obey some algebra which is also characteristic of the
quantum world).  The measures in the {\it conscious world} are given by a
subset of these amplitudes, the expectation values of a certain preferred set
of positive {\it awareness operators}.

In Sensible Quantum Mechanics, {\it probabilism}, or interpreting the
unconscious quantum world itself probabilistically, is an {\it aesthemamorphic
myth} (from the Greek $\alpha\iota\sigma\theta\eta\sigma\iota\sigma$:
perception, sense, sensation), rather analogous to the myth of animism that
ascribes living properties to inanimate objects.  It may be a convenient myth,
just as animism is a when we say such things as, ``Water seeks its own level,''
but it would give us a better understanding if we recognized it as a myth.
Thus probabilities don't (apply to) ``matter''; they are only in the ``mind.''

Also in SQM, the quantum state {\it never collapses}.  All sets of perceptions
with positive measure (given by the expectation values of different awareness
operators in the single unchanged Heisenberg state of the universe) actually
occur, so the theory is probabilistic not in the propensity sense but only in
the sense of ``frequencies'' (as ratios of the measures of the sets):  it is a
{\it many-perceptions} theory.

The measure of sets of perceptions in the conscious world permits a test in
principle of whether a perception is {\it typical}.  The Weak Anthropic
Principle can be generalized to the {\it Conditional Aesthemic Principle}:  our
conscious perceptions are likely to be typical perceptions in the conscious
world with its measure.

\section{Axioms of Sensible Quantum Mechanics}

\hspace{.25in}Sensible Quantum Mechanics (SQM) is given by the following three
fundamental postulates \cite{P95}:

 {\bf Quantum World Axiom}:  The unconscious ``quantum world'' $Q$ is
completely described by an appropriate algebra of operators and by a suitable
state $\sigma$ (a positive linear functional of the operators) giving the
expectation value $\langle O \rangle \equiv \sigma[O]$ of each operator $O$.

 {\bf Conscious World Axiom}:  The ``conscious world'' $M$, the set of all
perceptions $p$, has a fundamental measure $\mu(S)$ for each subset $S$ of $M$.

 {\bf Quantum-Consciousness Connection}:  The measure $\mu(S)$ for each set $S$
of conscious perceptions is given by the expectation value of a corresponding
``awareness operator'' $A(S)$, a positive-operator-valued (POV) measure
\cite{Dav}, in the state $\sigma$ of the quantum world:
 \begin{equation}
 \mu(S) = \langle A(S) \rangle \equiv \sigma[A(S)].
 \label{eq:1}
 \end{equation}

Here a perception $p$ is the entirety of a single conscious experience, all
that one is consciously aware of or consciously experiencing at one moment, the
total ``raw feel'' that one has at one time, or \cite{Lo} a ``phenomenal
perspective'' or ``maximal experience.''  [If a set of perceptions $S$ is
called a ``point of view'' (P.O.V.), then one may say that $A(S)$ is a POV for
a P.O.V.]

Since all sets $S$ of perceptions with $\mu(S) > 0$ really occur in SQM, it is
completely deterministic if the quantum state and the $A(S)$ are determined:
there are no random or truly probabilistic elements.  Nevertheless, because SQM
has measures for sets of perceptions, one can readily calculate ratios that can
be interpreted as conditional probabilities.  For example, one can consider the
set of perceptions $S_1$ in which there is a conscious memory of having tossed
a coin fifty times, and the set $S_2$ in which there is a conscious memory of
getting less than ten heads.  Then one can interpret
 \begin{equation}
 P(S_2|S_1)\equiv \mu(S_1\cap S_2)/\mu(S_1)
 \label{eq:2}
 \end{equation}
as the conditional probability that the perception is in the set $S_2$, given
that it is in the set $S_1$, that is, that a perception included a conscious
memory of getting less than ten heads, given that it included a conscious
memory of having tossed a coin fifty times.

In SQM the set $M$ of all perceptions is basic, and one can choose out of this
set any subset $S$ (by, e.g., the contents of the perceptions $p$ themselves,
or else by properties of the corresponding awareness operator $A(S)$ of the
subset $S$), but there is no absolutely preferred equivalence between
perceptions that could be used to classify them uniquely (except in {\it ad
hoc} ways) into sets corresponding to individual persons or minds.  Of course,
just as one can for objects such as ``chairs'' or ``protons'' that also
presumably do not have fundamental definitions in the ultimate theory of the
universe, one can make up {\it ad hoc} definitions that may be very good in
practice for classifying perceptions into persons or minds.  But that
classification is not fundamental to SQM.  As Hume \cite{Hume} wrote, ``what we
call a {\it mind}, is nothing but a heap or collection of different
perceptions, united together by certain relations, and suppos'd, tho' falsely,
to be endow'd with a perfect simplicity and identity.''  Since perceptions (or
what might crudely be called sensations) are basic to SQM, but not groupings of
them into minds, one might call SQM Mindless Sensationalism.

\section{Testing and Comparing SQM Theories}

\hspace{.25in}Since physics should be rooted in experience, we should have a
way to test and compare different candidate SQM theories.  If one had a theory
in which only a small subset of the set of all possible perceptions is
predicted to occur, one could simply check whether an experienced perception is
in that subset.  If it is not, that would be clear evidence against that
theory.  Unfortunately, in almost all SQM theories, almost all sets of
perceptions are predicted to have a positive measure, so these theories cannot
be excluded so simply.  For such many-perceptions theories, the best one can
hope for seems to be to find {\it likelihood} evidence for or against it.  Even
how to do this is not immediately obvious, since SQM theories merely give
measures for sets of perceptions rather than the existence probabilities for
any perceptions (unless the existence probabilities are considered to be unity
for all existing sets of perceptions, i.e., all those with nonzero measure, but
this is of little help, since almost all sets exist in this sense).

In order to test and compare SQM theories, it helps to hypothesize that the set
$M$ of all possible conscious
perceptions $p$ is a suitable topological space with a prior measure
 \begin{equation}
 \mu_0(S) = \int_{S}{d\mu_0(p)}.
 \label{eq:4}
 \end{equation}
Then, because of the linearity of positive-valued-operator measures over sets,
one can write each awareness operator as
 \begin{equation}
 A(S) = \int_S E(p)d\mu_0(p),
 \label{eq:5}
 \end{equation}
a generalized sum or integral of ``experience operators'' or ``perception
operators'' $E(p)$ for the individual perceptions $p$ in the set $S$.
Similarly, one can write the measure on a set of perceptions $S$ as
 \begin{equation}
 \mu(S) =  \langle A(S) \rangle = \int_S d\mu(p)
   = \int_S m(p) d\mu_0(p),
 \label{eq:6}
 \end{equation}
in terms of a measure density $m(p)$ that is the quantum expectation value of
the experience operator $E(p)$ for the same perception $p$:
 \begin{equation}
 m(p) = \langle E(p) \rangle \equiv \sigma[E(p)].
 \label{eq:7}
 \end{equation}

Now one can test the agreement of a particular SQM theory with a conscious
observation or perception $p$ by calculating the {\it typicality}  $T(p)$ that
the theory assigns to the perception:  Let $S_{\leq}(p)$ be the set of
perceptions $p'$ with $m(p') \leq m(p)$.  Then
 \begin{equation}
 T(p) \equiv \mu(S_{\leq}(p))/\mu(M).
 \label{eq:13}
 \end{equation}
For $p$ fixed and $\tilde{p}$ chosen randomly with the infinitesimal measure
$d\mu(\tilde{p})$, the probability that $T(\tilde{p})$ is less than or equal to
$T(p)$ is
 \begin{equation}
 P_T(p) \equiv P(T(\tilde{p})\leq T(p)) = T(p).
 \label{eq:14}
 \end{equation}
Thus the typicality $T_i(p)$ of a perception $p$ is the probability in a
particular SQM theory or hypothesis $H_i$ that another random perception will
have its typicality less than or equal to that of $p$ itself.  One can
interpret it as the likelihood of the perception $p$ in the particular theory
$H_i$, not for $p$ to exist, which is usually unity (interpreting all
perceptions $p$ with $m(p)>0$ as existing), but for $p$ to have a measure
density, and hence a typicality, no larger than it has.

Once the typicality $T_i(p)$ can be calculated for an experienced perception
assuming the theory $H_i$, one approach is to use it to rule out or falsify the
theory if the resulting typicality is too low.  Another approach is to assign
prior probabilities $P(H_i)$ to different theories (presumably neither
propensities nor frequencies but rather purely subjective probabilities,
perhaps one's guess for the ``propensities'' for God to create a universe
according to the various theories), say
 \begin{equation}
 P(H_i)=2^{-n_i},
 \label{eq:16}
 \end{equation}
where $n_i$ is the rank of $H_i$ in order of increasing complexity (my present
favorite choice for a countably infinite set of hypotheses if I could do this
ranking, which is another problem I will not further consider here).  Then one
can use Bayes' rule to calculate the posterior probability of the theory $H_i$
given the perception $p$ as
 \begin{equation}
 P(H_i|p)=\frac{P(H_i)T_i(p)}{\sum_{j}^{}{P(H_j)T_j(p)}}.
 \label{eq:15}
 \end{equation}

There is the potential technical problem that one might assign nonzero
prior probabilities to hypotheses $H_i$ in which the total measure $\mu(M)$ for
all perceptions is {\it not} finite, so that the right side of
Eq.~(\ref{eq:13}) may have both numerator and denominator infinite, which makes
the typicality $T_i(p)$ inherently ambiguous.  To avoid this problem, one might
use, instead of $T_i(p)$ in Eq.~(\ref{eq:15}), rather
 \begin{equation}
 T_i(p;S) = \mu_i(S_{\leq}(p)\cap S)/\mu_i(S)
 \label{eq:17}
 \end{equation}
for some set of perceptions $S$ containing $p$ that has $\mu_i(S)$ finite for
each hypothesis $H_i$.  This is related to a practical limitation anyway, since
one could presumably only hope to be able to compare the measure densities
$m(p)$ for some small set of perceptions rather similar to one's own, though it
is not clear in quantum cosmological theories that allow an infinite amount of
inflation how to get a finite measure even for a small set of perceptions
\cite{P95d}.  Unfortunately, even if one can get a finite measure by suitably
restricting the set $S$, this makes the resulting $P(H_i|p;S)$ depend on this
chosen $S$ as well as on the other postulated quantities such as $P(H_i)$.

\section{Properties of Experience Operators}

\hspace{.25in}Once one has a bare quantum theory (algebra of operators and
quantum state) for the quantum world, and the set $M$ of possible perceptions
$p$ with a prior measure for integrating any measure density $m(p)$ by
Eq.~(\ref{eq:6}) to get the corresponding measure $\mu(S)$ for sets $S$ of
perceptions $p$, the remaining feature of SQM to be determined is the
experience or perception operators $E(p)$, whose expectation values give the
measure density by Eq.~(\ref{eq:7}).  Assuming that the framework of SQM is
correct and that one knows what the set of possible perceptions is, the
uncertainty of the $E(p)$ encapsulates our ignorance of how the quantum world
produces conscious perceptions.

(One might object that even if we knew the full SQM theory with all the
$E(p)$'s, we would still not know {\it how} the quantum world produces
conscious perceptions.  This would be like saying that even if we have a law
for electromagnetism, we would still not know {\it how} a charged particle
produces an electromagnetic field.  But if we can say {\it what} perceptions or
{\it what} fields are produced in whatever various circumstances that may
occur, this is about as good an understanding as we can hope to get in physics,
though of course we would hope for the simplest description so that we can
describe as many things as possible with a small number of general principles.)

In \cite{P95} a large number of hypotheses were given (some compatible with
each other, but most incompatible alternatives) for the experience or
perception operators $E(p)$.  Here I shall not repeat most of them but shall
simply note that the the strongest one I am presently fairly comfortable with
(though without high confidence that it is true) is the Commuting Projection
Hypothesis of SQMPC:  $E(p)=P(p)$, a projection operator depending upon the
perception $p$, with $[P(p),P(p')]=0$ for all pairs of perceptions $p$ and
$p'$.  In many cases I would also think it might a plausibly good idealization
to make the Assumption of Perception Components, that each perception $p$
itself consists of a set of discrete components $c_i(p)$ contained within the
perception, say $p = \{c_i(p)\}$.  Then I would think that, at least as a
reasonably good approximation, one might strengthen the Commuting Projection
Hypothesis of SQMPC to the Commuting Product Projection Hypothesis of SQMPPC:
$E(p)=\prod_{i}{P[c_i(p)]}$, where each $P[c_i(p)]$ is a projection operator
that depends on the perception component $c_i(p)$, with {\it all} the
$P[c_i(p)]$'s commuting.

Then, although it is by no means required in SQM and indeed might be misleading
in circumstances in which these hypotheses do not hold, one might find it
heuristically advantageous to say that if the quantum state of the universe is
represented by the pure state $|\psi\rangle$, one can ascribe to the perception
$p$ the pure Everett ``relative state''
 \begin{equation}
 |p\rangle=\frac{E(p)|\psi\rangle}{\parallel  E(p)|\psi\rangle\parallel}
 =\frac{E(p)|\psi\rangle}
 {\langle\psi|E(p)E(p)|\psi\rangle^{1/2}}.
 \label{eq:P3}
 \end{equation}
Alternatively, if the quantum state of the universe is represented by the
density matrix $\rho$, one can associate the perception with a relative density
matrix
 \begin{equation}
 \rho_p=\frac{E(p)\rho E(p)}{Tr[E(p)\rho E(p)]}.
 \label{eq:P4}
 \end{equation}
Then if one is willing to say that $m(p)=Tr[E(p)\rho]$ is the absolute
probability for the perception $p$ (which might seem natural at least when
$E(p)$ is a projection operator, though I am certainly not advocating this
na\"{\i}ve interpretation), one might also na\"{\i}vely interpret
$Tr[E(p')\rho_p]$ as the conditional probability of the perception $p'$ given
the perception $p$.

 Another thing one can do with two perceptions $p$ and $p'$ is to calculate an
``overlap fraction'' between them as
 \begin{equation}
 f(p,p')=\frac{\langle E(p)E(p')
 \rangle\langle E(p')E(p)\rangle}
 {\langle E(p)E(p)\rangle\langle E(p')E(p')\rangle}.
 \label{eq:P5}
 \end{equation}
If the quantum state of the universe is pure, this is the same as the overlap
probability between the two Everett relative states corresponding to the
perceptions:  $f(p,p')=|\langle p|p'\rangle|^2$.  Thus one might in some sense
say that if $f(p,p')$ is near unity, the two perceptions are in nearly the same
one of the Everett ``many worlds,'' but if $f(p,p')$ is near zero, the two
perceptions are in nearly orthogonal different worlds.  However, this is just a
manner of speaking, since I do not wish to say that the quantum state of the
universe is really divided up into many different worlds.  In a slightly
different way of putting it, one might also propose that $f(p,p')$, instead of
$Tr[E(p')\rho_p]$, be interpreted as the conditional probability of the
perception $p'$ given the perception $p$.  Still, I do not see any evidence
that $f(p,p')$ should be interpreted as a fundamental element of Sensible
Quantum Mechanics.  In any case, one can be conscious only of a single
perception at once, so there is no way in principle that one can test any
properties of joint perceptions such as $f(p,p')$.

\section{Perceptions of Schr\"{o}dinger's Cat}

\hspace{.25in}The framework of Sensible quantum mechanics allows one to discuss
questions of what one would be perceived in the experiment of Schr\"{o}dinger's
cat \cite{Sch}, and a detailed SQM theory would answer such questions.  One
such question is whether one could directly perceive a superposition of, say,
alive plus dead.

If perceptions are of alive ($A$) versus dead ($D$), so $\langle E(p_{\rm
alive})\rangle\propto\langle |A\rangle\langle A|\rangle$ and $\langle E(p_{\rm
dead})\rangle\propto\langle |D\rangle\langle D|\rangle$, then in an appropriate
basis for this the quantum state resulting from the idealized Schr\"{o}dinger's
cat experiment may have the form
 \begin{equation}
 |\psi\rangle\propto |A\rangle_{\rm head}
 |A\rangle_{\rm body}|A\rangle_{\rm rest} +
 |D\rangle_{\rm head}|D\rangle_{\rm body}
 |D\rangle_{\rm rest} + (\mbox{other terms}),
 \label{eq:47}
 \end{equation}
where I have conceptually divided the system (e.g., the universe) into the head
of the cat, the body of the cat, and the rest of the universe.  Here ``(other
terms)'' denotes other components of the quantum state in which the
Schr\"{o}dinger's cat experiment has not been done and there is no perception
that it has; I shall here ignore the other irrelevant perceptions whose measure
arises from such terms.  One now finds that in this idealized state there is an
equal nonzero measure density for the perceptions $p_{\mbox{head alive, body
alive,\ldots}}$ and $p_{\mbox{head dead, body dead,\ldots}}$, but no measure
density for $p_{\mbox{head alive, body dead,\ldots}}$ or $p_{\mbox{head dead,
body alive,\ldots}}$.  In other words, there is not a unique perception of
whether the cat is alive or dead, but in each perception that the experiment
has been done (given by part of the unspecified perception components denoted
by the \ldots in the subscripts for the $p$'s), there is perfect agreement that
the head and body are either both alive or both dead.

If, on the other hand, perceptions were of the linear combinations
$|+\rangle\propto |A\rangle + |D\rangle$ and $|-\rangle\propto |A\rangle -
|D\rangle$, so $\langle E(p_+)\rangle\propto\langle |+\rangle\langle +|\rangle$
and $\langle E(p_-)\rangle\propto\langle |-\rangle\langle -|\rangle$, then in
the appropriate basis for this the quantum state has the form
 \begin{eqnarray}
 |\psi\rangle\propto |+\rangle_{\rm head}
 |+\rangle_{\rm body}|+\rangle_{\rm rest} +
 |-\rangle_{\rm head}|-\rangle_{\rm body}
 |+\rangle_{\rm rest} +\nonumber \\ |+\rangle_{\rm head}
 |-\rangle_{\rm body}|-\rangle_{\rm rest} +
 |-\rangle_{\rm head}|+\rangle_{\rm body}
 |-\rangle_{\rm rest} + (\mbox{other terms}).
 \label{eq:48}
 \end{eqnarray}
In this case one one gets equal measure densities for all four possible
perceptions $p_{++}$, $p_{--}$, $p_{+-}$, and $p_{-+}$, so only $2^{-1}$ of the
perceptions agree about the head and body.

If one had instead conceptually divided the cat into $n>2$ parts which were all
either dead or alive, then if perception components were of that property (dead
or alive), there would be, in the idealized Schr\"{o}dinger's cat experiment,
just as in the $n=2$ case above, no single perception of whether the cat or any
of its parts were dead or alive, but within each perception there would be
total agreement between the perception components of whether each part of the
cat were dead or alive, so there would be no confusion within any single
multi-component perception.  On the other hand, if perception components were
of the $+$ and $-$ properties of each part of the cat, all possible $2^n$
orders would occur with equal weight, so only a fraction of $2^{1-n}$ of the
perceptions would have agreement for all $n$ parts of the cat., and in the
remaining bulk of the perceptions there would be confusion as to whether the
entire cat is dead or alive.

Note that without knowing what the experience operators actually are, this
analysis cannot answer the question of how each part of the cat is perceived.
However, it does show that {\it if} the perception is to give agreement between
the components corresponding to the different parts of the cat, and if (as in
the idealized Schr\"{o}dinger's cat experiment) there is a complete correlation
between the parts as to whether they are alive or dead, {\it then} the
perception components should be of whether the corresponding part of the cat is
alive or dead, rather than being of the linear combinations $+$ or $-$.  This
does seem to fit our experience much better than the other possibility, so
empirically we can say that we tend to have relatively unconfused perceptions,
at least compared to the maximum confusion conceivable.  There is still the
mystery of {\it why} this is so, and I am tempted to paraphrase Einstein to
say, ``The most confusing thing about perceptions is that they are not
generally confusing.''

One possible attempt at an explanation is to argue that if our perceptions were
confused, we would not respond coherently to our environment and so would not
survive.  However, this assumes that our perceptions really do affect our
actions (e.g., part of the quantum state) rather than just being passively
produced by the quantum state as epiphenomena.  SQM describes only the
production of perceptions (i.e., the determination of their measure) by the
quantum state and the awareness or experience operators but does not describe
any action of the perceptions back on the state.  If the state were really
unaffected by the perceptions, and if the survival of organisms can be
described by the properties of the state, then this survival would be totally
unaffected by the perceptions, and, in particular, by whether they are confused
or unconfused.  Of course, there may be some other explanation of the coherence
of perceptions besides the survival value for an organism, but the
attractiveness of that particular explanation does at least suggest that
perceptions do have an action back on the quantum state.  Another suggestive
argument for the same conclusion, and a sketch of how this back reaction might
fit in with the presently known laws of physics, will be given near the end of
this paper.

\section{Location and Time of Perceptions in QFT}

\hspace{.25in}In general, perceptions $p$ are associated with experience
operators $E(p)$ (or sets of perceptions with awareness operators $A(S)$), but
not with times or locations.  Instead, Sensible Quantum Mechanics transcends
quantum theories in which space and time are fundamental.

But for quantum field theory (QFT) in a classical curved globally hyperbolic
background spacetime without symmetries (so that points can be uniquely
identified by the background), one can make an {\it ad hoc} definition of a
time and location associated with a perception $p$:  Choose a one-parameter
(time $t$) sequence of Cauchy hypersurfaces.  For each time $t$, point $P$ on
the hypersurface, and radius $r$, let $E'(p;t,P,r)$ be $E(p)$ written in terms
of the fields and conjugate momenta at $t$ but with terms at distances greater
than $r$ from the point $P$ truncated.  Define
 \begin{equation}
 F(p;t,P,r)=\frac{\langle E(p)E'(p;t,P,r)\rangle
 \langle E'(p;t,P,r)E(p)\rangle}
 {\langle E(p)E(p)\rangle
 \langle E'(p;t,P,r)E'(p;t,P,r)\rangle},
 \label{eq:P6}
 \end{equation}
 \begin{equation}
 r(p,t,P) = \min{[r: F(p;t,P,r) = 1/2]},
 \label{eq:P7}
 \end{equation}
 \begin{equation}
 r(p;t)=\min{[r(p;t,P): P\mbox{ on hypersurface of time }t]},
 \label{eq:P8}
 \end{equation}
 \begin{equation}
 P(p;t) = P \mbox{ such that } F(p;t,P,r(p;t)) = 1/2,
 \label{eq:P9}
 \end{equation}
 \begin{equation}
 t_p = t \mbox{ such that } r(p;t) = \min{[r(p;t'): t']},
 \label{eq:P10}
 \end{equation}
 \begin{equation}
 r_p = r(p;t_p) = \min{[r(p;t):t]},
 \label{eq:P11}
 \end{equation}
 \begin{equation}
 P_p = P(p;t_p).
 \label{eq:P12}
 \end{equation}
This locates the perception as crudely occurring mostly within the smallest
possible ball of geodesic radius $r_p$ from the point $P_p$ on the hypersurface
at time $t_p$.

\section{Questions and Speculations}

\hspace{.25in}One can use the framework of Sensible Quantum Mechanics to ask
questions and make speculations that might be difficult otherwise:

1.  What regions (presumably inside brains) are most responsible for
perceptions?

2.  Does the region depend significantly on the character of the perception?

3.  Can one have two quite different perceptions, $p$ and $p'$, with
$f(p,p')\approx 1$ (i.e., in nearly the ``same Everett world''), $t_p = t_{p'}$
(i.e., most localized at the same times), and with both $P_p$ and $P_{p'}$ in
the same brain?  We generally believe this is possible for two different
brains, but can one single brain have two different perceptions (and not just
two different components of a single perception) at once?

4.  If so, can the two balls of radius $r_p$ and $r_{p'}$ overlap?  (I.e., can
the same region of the brain have two different perceptions at once?)

5.  How does the measure density $m(p)$ depend on brain characteristics?

6.  Is it correlated with intelligence, so that in some sense brighter brains
give a larger measure of perceptions?

7.  Does this explain why you perceive yourself as human rather than insect,
although there are many more insects than humans?

8.  Does this explain why you may perceive yourself as more intelligent than
most other people?  (In other words, are you more typical, when weighted by the
measure for perceptions, than you might otherwise have thought ?)

9.  Can electronic computers give significant measures for perceptions?

10.  Are human brains more efficient than most electronic computers, no matter
how intelligent (at least in information-processing capabilities), in producing
conscious perceptions?  (If so, the measure for perceptions would not be
correlated purely with intelligence in this sense.)

11.  Does this explain why you do not perceive yourself as one of trillions of
self-replicating computers that might colonize the Galaxy?

12.  Over what region of spacetime do human brains have the dominant measure of
conscious perceptions in our Everett world $\rho_p$?

The Conditional Aesthemic Principle would predict that our conscious
perceptions are likely to be typical perceptions in the conscious world with
its measure.  Thus it would predict that it is unlikely that the overwhelming
bulk of conscious perceptions in the universe would have a measure density (and
hence a typicality) larger than ours, though of course it allows for the
possibility that many other types of perceptions could have comparable measure
densities and typicalities, so it does not predict that the dominant
perceptions should be peculiarly human.

\section{An Analogy for the Mind-Body Problem}

\hspace{.25in}To explain the mind-body relation in Sensible Quantum mechanics
in terms of an analogy, consider a classical model of spinless massive point
charged particles and an electromagnetic field in Minkowski spacetime.  The
charged particles can be considered to be analogous to the quantum world (or
the quantum state part of it), and the electromagnetic field can be considered
to be analogous to the conscious world (the set of perceptions with its measure
$\mu(S)$).

At the level of a simplistic materialist mind-body philosophy, one might merely
say that the electromagnetic field is part of, or perhaps a property of, the
material particles.  One cannot say for certain that this is wrong, but it does
not lead to much understanding of the electromagnetic field merely to say that.
 Similarly, one cannot rule out the claim that consciousness is merely a
property of the quantum world, but just saying that does not give much insight
into consciousness.

At the level of Sensible Quantum Mechanics, the charged particle worldlines are
the analogue of the quantum state, the retarded electromagnetic field
propagator (Coulomb's law in the nonrelativistic approximation) is the analogue
of the awareness operators, and the electromagnetic field determined by the
worldlines of the charged particles
and by the retarded propagator is the analogue of the conscious world.  (Here
one can see that this analogue of Sensible Quantum Mechanics is valid only if
there is no free
incoming electromagnetic radiation.)

One might propose an extension of Sensible Quantum Mechanics, say Sensational
Quantum Mechanics, in which the conscious world may affect the quantum world.
The analogue of this would be the case in which the charged particle worldlines
are partially determined by the electromagnetic field through the change in the
action it causes.  (This more unified framework better explains the previous
level but does not violate its description, which
simply had the particle worldlines given.)

(A motivation for considering the possibility that the conscious world might
have an effect on the quantum world, besides the actual effect in the
electromagnetic analogue being considered here, was given at the end of the
section above on the experiment of Schr\"{o}dinger's cat.  Another motivation
which occurred to me earlier would be the desire to explain the correlation
between will and action, i.e., why I feel I do as I please.  An easy way to
circumvent the objection that such an effect would violate the known laws of
physics, in particular those of energy-momentum conservation, would be to have
desires in the conscious world affect, in a coordinate-invariant way that would
thus preserve energy-momentum conservation, the action functional that is used
in a path integral giving the quantum state.)

At a yet higher level in the analogue, there is the possibility of incoming
free electromagnetic waves, which would violate the previous frameworks that
assumed the electromagnetic field was uniquely determined by the charged
particle worldlines.  An analogous suggestion for
intrinsic degrees of freedom for consciousness has been made by Linde
\cite{Lin90}.

Finally, at a still higher level, there might be an even more
unifying framework in which both charged particles and the electromagnetic
field are seen as modes of a single entity (e.g., to take a popular current
speculation, a superstring).  Such a more unified framework for the mind-body
problem might exist as well, but I suspect that one will not get to such a
framework with any significant content without examining the lower levels first
and then hopefully finding a complete unified description from which the lower
levels can be shown to emerge by some sort of reduction or approximation.

Thus Sensible Quantum Mechanics may be only a framework for the next step in
understanding the relation between mind and body or between conscious
observations and quantum mechanics.  However, it does seem to give a glimpse of
a yet-to-be-completed quantum theory that, when filled in in detail, could not
be criticized for being inherently incomplete in not predicting precisely what
happens during observations.

I perceive that my thoughts on this subject have benefited by interactions with
many people listed in my more complete account in \cite{P95}, but here I
especially want to acknowledge the continued e-mail interaction with my
previous co-author Shelly Goldstein \cite{GP} (whom I shall first meet in
person only after writing these words).  Financial support has been provided by
the Natural Sciences and Engineering Research Council of Canada.



\baselineskip 7pt

\begin{thebibliography}{99}

\bibitem{E} H. Everett, III, Rev.\ Mod.\ Phys. {\bf 29}, 454 (1957); B. S.
DeWitt and N. Graham, eds., {\em The Many-Worlds Interpretation of Quantum
Mechanics} (Princeton University Press, Princeton, 1973).

\bibitem{P94a} D. N. Page, ``Probabilities Don't Matter,'' to be published in
{\em Proceedings of the 7th Marcel Grossmann Meeting on General Relativity},
edited by M. Keiser and R. T. Jantzen (World Scientific, Singapore 1995)
(University of Alberta report Alberta-Thy-28-94, Nov. 25, 1994), gr-qc/9411004.

\bibitem{P94b} D. N. Page, ``Information Loss in Black Holes and/or Conscious
Beings?'' to be published
in {\em Heat Kernel Techniques and Quantum Gravity}, edited by S. A. Fulling
(Discourses in Mathematics and Its Applications, No. 4, Texas A\&M University
Department of Mathematics, College Station, Texas, 1995) (University of Alberta
report Alberta-Thy-36-94, Nov. 25, 1994), hep-th/9411193.

\bibitem{P95} D. N. Page, ``Sensible Quantum Mechanics:  Are Only Perceptions
Probabilistic?'' (University of Alberta
report Alberta-Thy-05-95, June 7, 1995), quant-ph/9506010.

\bibitem{P95b} D. N. Page, ``Attaching Theories of Consciousness to Bohmian
Quantum Mechanics,'' to be published in {\em Bohmian Quantum Mechanics and
Quantum Theory:  An Appraisal}, edited by J. T. Cushing, A. Fine, and S.
Goldstein (Kluwer, Dordrecht, 1996) (University of Alberta report
Alberta-Thy-12-95, June 30, 1995), quant-ph/9507006.

\bibitem{P95d} D. N. Page, ``Aspects of Quantum Cosmology,'' to be published in
{\em Proceedings of the International School of Astrophysics ``D. Chalonge,''
4th Course:  String Gravity and Physics at the Planck Energy Scale, Erice,
Sicily, 8-19 September 1995}, edited by N. Sanchez (Kluwer, Dordrecht, 1996)
(University of Alberta report Alberta-Thy-14-95, July 10, 1995), gr-qc/9507025.

\bibitem{Dav} E. B. Davies, {\em Quantum Theory of Open Systems} (Academic
Press, London, 1976).

\bibitem{Lo} M. Lockwood, {\em Mind, Brain and the Quantum:  The Compound `I'}
(Basil Blackwell, Oxford, 1989).

\bibitem{Hume} D. Hume, {\em A Treatise of Human Nature}, reprinted from the
original edition in three volumes and edited by L. A. Selby-Bigge (Clarendon,
Oxford, 1888), p. 207.

\bibitem{Sch} E. Schr\"{o}dinger, Naturwissenschaften {\bf 23}, 807 (1935),
English translation by J. D. Trimmer, Proc. Am.\ Phil.\ Soc.\ {\bf 124}, 323
(1980), reprinted in {\em Quantum Theory and Measurement}, edited by J. A.
Wheeler and W. H. Zurek (Princeton University Press, Princeton, 1983), p.~152.

\bibitem{Lin90} A. Linde, {\it Particle Physics and Inflationary Cosmology}
(Harwood Academic Publishers, Chur, Switzerland, 1990), p.~317.

\bibitem{GP} S. Goldstein and D. N. Page, Phys.\ Rev.\ Lett.\ {\bf 74}, 3715
(1995), gr-qc/9403055.

\end{thebibliography}
\end{document}